\documentclass[runningheads]{llncs}
\usepackage[T1]{fontenc}
\usepackage{graphicx}

\begin{document}
\title{Towards Modeling Cybersecurity Behavior of Humans in Organizations}
\titlerunning{Security and Human Behavior in Organisations}
\author{Klaas Ole Kürtz\orcidID{0009-0002-5631-2444}}
\authorrunning{K. O. Kürtz}
\institute{Kiel University of Applied Sciences, Sokratesplatz 2, 24149 Kiel, Germany \email{klaas.ole.kuertz@haw-kiel.de}}
\maketitle

\begin{abstract}
We undertake a comprehensive and structured synthesis of the drivers of human behavior in cybersecurity, focusing specifically on people within organizations (i.e., especially employees in companies), and integrate key concepts such as awareness, security culture, and usability into a coherent theoretical framework. This model is then compared with several relevant behavioral models that fundamentally represent drivers of human behavior.

Additionally, we discuss how this theoretical framework can help the domain of agentic AI security: We argue that as AI systems increasingly act as autonomous agents within organizations and based on natural language processing, they also exhibit vulnerabilities analogous to human behavioral risks. Consequently, we propose that this human-centric model offers a blueprint for developing additional security strategies against manipulation attacks targeting AI agents.
\end{abstract}

\section{Introduction and methodology}

Despite technological advances and active research, cybersecurity remains a challenge. Humans are a central factor in cybersecurity~\cite{berensHumanFactorsSecurity2021,cybok-hf} as is illustrated by current developments adressing the human factor, such as the use of generative artificial intelligence (AI) to scale and professionalize social engineering~\cite{schmittDigitalDeceptionGenerative2024}, or attacks via channels that circumvent organizations' technical safeguards (e.g., contact via professional social networks)~\cite{SaranyaAIPoweredPhishingDetection2025}.

This reveals a dualism of the human factor in cybersecurity: On the one hand, humans are a critical attack surface and one of the largest attack vectors~\cite{maalemlahcenReviewInsightBehavioral2020}. On the other hand, under the right conditions, human intelligence can become the most adaptable and resilient element of a cybersecurity strategy (e.g., referred to as a ``human firewall''~\cite{joergensHumanFirewall2023}). Shifting the focus from blaming the user to developing a resilient system that considers human cognition and organizational realities represents a crucial challenge for modern security strategies.

At the same time, artificial intelligence systems are changing cybersecurity in many ways, also influencing how humans behave or should behave: First, AI itself can be a tool in attacks (e.g., social engineering, which puts more emphasis on human behavior in detecting such attacks); second, it can also be a tool in defense (e.g., detecting anomalies also in how e.g. employees behave); third, AI systems themselves are the target of attacks (e.g., prompt injection targeting the human language component of AI systems); and fourth, the increasing use of AI is changing the overall IT landscape (e.g., the increase in AI-generated code), which will have an overall impact on cybersecurity and the role of human behavior in organizations.

Our contribution therefore consists of a practical, structuring synthesis of the factors of human behavior in cybersecurity, particularly in organizations, and a comparison with relevant existing theories.

To derive a holistic model relevant to professional environments, we employed the following synthesis approach. We first conducted a review of established academic behavioral science theories to identify fundamental psychological drivers of human action, such as Protection Motivation Theory~\cite{PMT} and Theory of Planned Behavior~\cite{TPB}). This theoretical foundation was then cross-referenced with scientific and applied cybersecurity management literature and qualitative data, including insights from informal expert consultations and published perspectives of Chief Information Security Officers (CISOs). This methodology allowed us to filter abstract psychological concepts through the lens of practical organizational reality, ensuring the resulting framework captures both the internal cognitive processes of employees and the external systemic pressures of the workplace.

\section{Structuring synthesis of factors of human behavior in relation to cybersecurity in organizations}\label{section:driver-human-beahvior}

The interdisciplinary research field surrounding the role of humans in cybersecurity has produced a wealth of insights in recent decades (e.g., ``Security and Human Behavior''~\cite{STAST2019,STAST2021,dhillonInformationSystemsSecurity2021,STAST2022}, ``Usable Security''~\cite{markottenUserCenteredSecurityEngineering2002,payneBriefIntroductionUsable2008,SOUPS2023,SOUPS2024,SOUPS2025}).

\begin{figure}[t]
	\includegraphics[width=1.00\textwidth]{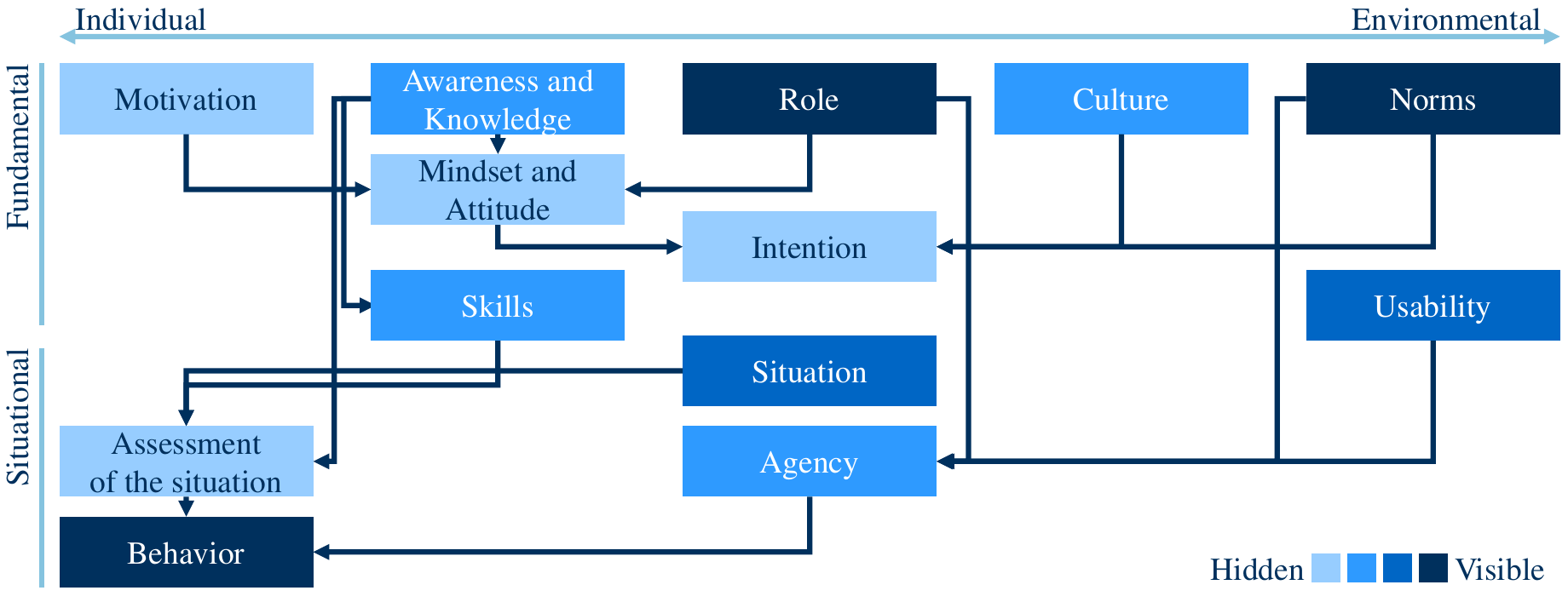}
	\caption{Factors of Human Behavior in Relation to Cybersecurity in Organizations}
	\label{fig:driver-human-behavior}
\end{figure}

The focus of our work is the behavior of users and the causes of this behavior in relation to cybersecurity-relevant situations, primarily in organizations (e.g., companies), and thus goes partly beyond existing models that primarily focus on the individual decisions of users~\cite{addaeExploringUserBehavioral2019,maalemlahcenReviewInsightBehavioral2020,schalteggerHumanBehaviorCybersecurity2025}. For our analysis, we structure the research field in Figure~\ref{fig:driver-human-behavior} with the aim of synthesizing the essential factors in terms of influencing factors or drivers and representing their interrelationship in the form of a practical model, on the basis of which we can later draw analogies to AI agents and then build a research agenda for the security of AI agents. We deconstruct the factors that precede and shape every security-relevant action and arrange them into a logical, causal flow that encompasses both the individual's internal world and the external environment in which they operate. The essential relationships between the main factors are also listed; however, these relationships and the logical flow necessarily represent a significant simplification. Numerous further gradations and cross-relationships exist between the factors: Human decision-making is more complex than the model presented here or the behavioral science models used.

Our model uses three dimensions to classify the main factors, the scale of which is only conceptual:

\begin{itemize}
\item Logical flow from fundamental to situational factors: Fundamental factors are shaped or predetermined rather independently of specific situations (e.g., the basic culture of an organization and the role of an individual) and then influence the assessment or behavior in a specific situation.

\item Comparison of individual and environmental factors: Individual, internal factors relate to the individual person, while environmental factors concern, for example, an entire organization (e.g., a company) or social group. An example of this is organizational culture in relation to cybersecurity, which is primarily an external environmental factor (although it interacts with the individual), and in contrast, for example, motivation, which is primarily individual (although influenced by the environment).

\item Comparison of hidden and visible factors: Hidden elements that precede the actually measurable or observable elements, such as external norms or a person's actual behavior in a specific situation, influence this behavior -- such as individual motivation or behavioral intention, which are more difficult to measure or observe.
\end{itemize}

Our model is shown in Figure~\ref{fig:driver-human-behavior} and contains the following main factors with further associated factors:

\begin{itemize}

\item \textbf{Motivation}: Individual motivation or drive, both intrinsic and extrinsic (e.g., belonging to a social group, helpfulness, financial interests or incentives), as well as feelings (e.g., fear); compare~\cite{faganFollowNotFollow2018}

\item \textbf{Awareness and Knowledge} about cybersecurity including errors contained therein (e.g., cognitive distortions, biases); compare~\cite{zwillingCyberSecurityAwareness2022}

\begin{itemize}
\item Awareness and knowledge of threats (abstract and concrete) and individual threat assessment (perceived vulnerability to threats and their perceived severity)

\item Awareness and knowledge of countermeasures and individual effectiveness assessment (self-efficacy expectation, i.e., perceived effectiveness of one's own actions)
\end{itemize}

\item \textbf{Role} regarding cybersecurity: Where applicable, a formalized role (e.g., implicit and explicit responsibilities) in the sense of a division of tasks (e.g., employees in the IT security department, security champion in a non-IT department); compare~\cite{granovaChangingHeartsMinds2023}

\item \textbf{Mindset and Attitude} towards cybersecurity: Individual perspective on the topic and one's own role (e.g., engaging or ignoring, also security fatigue) -- based in particular on individual motivation, knowledge and role; compare~\cite{schoenmakersSecurityMindsetCharacteristics2023}

\item \textbf{Culture} in the organization or social group with regard to cybersecurity (implicit behavior, e.g., leadership, security and error culture in companies, visibility of role models, knowledge sharing, perceived coercion or voluntariness); compare~\cite{corradiniBuildingCybersecurityCulture2020,uchenduDevelopingCyberSecurity2021,mwimSystematicReviewFactors2022,enisaCybersecurityCultureGuidelines2024}

\item \textbf{Norms} and Regulations, e.g., relevant laws or explicit guidelines, also relating to organizations, as well as perceived behavioral control regarding such norms (e.g., density of control and visibility of actual behavior); compare~\cite{gisladottirResilienceCyberSystems2017}

\item \textbf{Intention}: individual behavioral intention (e.g., intended vigilance or engagement); based in particular on individual mindset and attitude, culture, and norms; compare~\cite{gratianCorrelatingHumanTraits2018,ceranWhatInfluencesHuman2025}

\item \textbf{Skills}: Individual competence to act (e.g., competence to monitor anomalies unconsciously, to recognize attack situations, to apply adversarial thinking, to intuitively evaluate risks, to chose an adequate response for prevention, detection, or reaction to attacks); also based on awareness and knowledge; compare~\cite{hoUnderstandingEfficacyPhishing2025}

\item \textbf{Usability} (also User Friendliness and User Experience): Usability and user experience of the security-related correct use of IT systems (e.g., security by default, security friction, salience of security-relevant information); compare~\cite{ebertWhenInformationSecurity2023}

\item \textbf{Agency}: Individually perceived autonomy and authority to act in the specific situation (e.g., existence of decision-making latitude in response to situations) or self-efficacy; based in particular on norms, role, and usability; compare~\cite{heSelfEfficacyVariableBehavioral2014,borgertSelfEfficacySecurityBehavior2024}

\item \textbf{Situation}: Situational factors in an actual attack situation, both individual (e.g., cognitive load, attention and fatigue, stress) and environmental in the organization (e.g., external time pressure, technical environment); compare~\cite{vishwanathWhyPeopleGet2011}

\item \textbf{Assessment of the Situation}: The sum of conscious and unconscious assessments of an actual attack situation, e.g., significance and relevance, degree of activation or alertness –- based on the situation, awareness and knowledge, and skills; compare~\cite{vanschaikRiskPerceptionsCybersecurity2017,kostyukMicrofoundationsStateCybersecurity2021}

\item \textbf{Behavior}: Actual behavior in a given situation with regard to cybersecurity -- based in particular on the individual's assessment of the situation and their agency to act; includes both conscious action (preceded by an active decision) and unconscious, intuitive action.
\end{itemize}

\section{Comparison with relevant behavioral science theories}\label{section:comparison}

Various theories, particularly from behavioral psychology, offer explanatory models with different, complementary perspectives for the complex decision-making processes that lead to security-relevant behavior. To verify the completeness of our model, we are comparing it with relevant existing models.

We draw on the most relevant established models -- and assign the elements or factors of the corresponding theories to the main factors in our model by referring to them in italics.

\subsection{Cognitive or rational theories}

\begin{itemize}
\item The Theory of Planned Behavior (TPB)~\cite{TPB} postulates that a person's behavior is determined by their \emph{behavioral intention}, which in turn is influenced by their \emph{attitude} towards the behavior, \emph{norms}, and perceived behavioral control (included in \emph{norms} in the model).

\item The Protection Motivation Theory (PMT)~\cite{PMT} explains how individuals respond to threats based on their assessment of the threat, such as severity and vulnerability (modeled through \emph{awareness and knowledge}), and their ability to respond (modeled through \emph{skills} and \emph{agency}). PMT emphasizes the role of fear appeals but also cautions that excessive fear without adequate coping mechanisms can lead to avoidance behavior. When people perceive a threat as severe and likely but do not believe they can do anything effective about it (low agency and self-efficacy), they may fall into a state of resignation.

\item The Technology Acceptance Model (TAM)~\cite{TAMa,TAMb} focuses on the acceptance and use of new technologies, but can also be applied to analyze the acceptance of security measures. It states that perceived usefulness (modeled through \emph{awareness and knowledge} as well as \emph{skills}) and perceived ease of use (modeled through \emph{usability}) influence attitudes towards technology and thus the intention to use it (\emph{intention}, \emph{behavior})

\item The Unified Theory of Acceptance and Use of Technology (UTAUT)~\cite{UTAUT}  extends the Technology Acceptance Model and integrates elements from several theories. It identifies performance expectancy (\emph{awareness and knowledge}, \emph{skills}, \emph{agency}), effort expectancy (\emph{usability}), social influence (\emph{culture}), and facilitating conditions (e.g., \emph{usability}) as determinants of behavioral intention and actual usage; additionally, it utilizes moderating social factors (such as gender and age) not considered here.

\item The Dual Process Theory, in Kahneman's variant~\cite{DualProcessKahnemann}, postulates that human information processing occurs in two ways: a fast, intuitive System 1 and a slower, more analytical System 2. While security training often aims to activate System 2 \emph{awareness and knowledge} as well as \emph{skills}), attacks such as social engineering often attempt to have System 1 determine behavior through factors in their attack (\emph{situation}).

\item Self-Determination Theory (SDT)~\cite{SDT} focuses on intrinsic motivation and states that people are most motivated when they experience autonomy (\emph{role}, \emph{agency}), competence (\emph{awareness and knowledge} as well as \emph{skills}) and social relatedness (\emph{culture}).
\end{itemize}

\subsection{Contextual, organizational or process theories}

\begin{itemize}
\item The Fogg Behavior Model describes the reasons why users perform a particular action or not. Behavior occurs precisely when users want to adopt the behavior (\emph{motivation}, \emph{intention}), are easily able to do so (\emph{skills}, \emph{usability}), and something triggers the action (\emph{situation}).

\item The Deterrence Theory~\cite{DeterrenceTheoryB,DeterrenceTheoryA} states that undesirable behavior of individuals (in this context, e.g., non-compliance with security guidelines) can be reduced by punishment, with the strength of the effect being influenced by the perceived severity of the punishment, the probability of detection, and the speed of the response (\emph{culture} and \emph{norms}).

\item The OODA loop~\cite{OODA} describes a decision loop, originally developed in the military field, a reaction to an event consisting of the four phases Observe (\emph{awareness and knowledge}, \emph{situation}), Orient (in particular \emph{mindset and attitude}), Decide (\emph{intention}) and Act (\emph{behavior}).

\item The ``Four Building Blocks of Change''~\cite{InfluenceModel} are an example of a practice-oriented model, identifying four essential building blocks for change in organizations, for example for establishing a sustainable security culture: A compelling narrative (\emph{awareness and knowledge}), strengthening of \emph{skills}, formal mechanisms (\emph{role}, \emph{norms}) and role models (\emph{culture}).
\end{itemize}

While these established theories provide robust mechanisms for explaining specific internal cognitive processes (such as the threat appraisal in PMT or intention in TPB) or technology acceptance (TAM), they predominantly view the individual in isolation. Our proposed framework distinguishes itself by explicitly integrating these internal psychological drivers with external organizational realities---specifically culture, role, and norms---and dynamic situational factors relevant for ad-hoc decisions in attack situations. By capturing the interplay between an employee's internal state (individual) and the workplace environment (environmental), our model addresses the specific sociotechnical complexity of cybersecurity that purely individualistic behavioral models often do not capture.

\section{Outlook and future work on the cybersecurity of agentic AI systems}\label{section:research-agenda}

The systematic model of these drivers aims to help form a foundation for two intended areas of interdisciplinary research, taking elements from behavioral science, sociology, economic psychology, risk research, and other disciplines, applying them to cybersecurity.

First, we argue future research can shift focus from isolated individual actions to behavior embedded within the organizational context: Our model highlights that an individual's security behavior is not solely a product of internal factors like motivation or skills, but is heavily moderated by environmental drivers such as organizational culture, formal roles, and norms including the perceived control on those norms. We propose investigating how these external organizational structures specifically suppress or amplify individual traits. For instance, understanding to what extent a strong ``security culture'' can compensate for low individual ``security awareness'' is critical for designing more resilient organizational defenses that go beyond training the individual.

Secondly, we argue that the security of agentic AI systems is no longer a purely technical challenge, but shows strong parallels to aspects of human behavior in cybersecurity: Agentic AI systems or AI agents are increasingly expected to make independent decisions and perform active actions on behalf of users~\cite{acharyaAgenticAIAutonomous2025,hosseiniRoleAgenticAI2025,xiRisePotentialLarge2025,sapkotaAIAgentsVs2026}. This increasing autonomy and human-like interaction means that agentic AI systems are evolving into independent actors in the cybersecurity sense and can exhibit their own classes of quasi-human vulnerabilities: Attack techniques such as prompt injection are targeting AI models' ``understanding'' of the input and the decision-making levels of AI models, to exploit their training based on human language.

Consider the following attack scenario as an example: An AI agent conducts web research on behalf of a user, meaning the AI agent accesses external sources on the web. One of these sources contains hidden malicious instructions, unbeknownst to the user or the AI agent. These instructions could, for example, aim to violate confidentiality (disclosure of confidential user information) or integrity (manipulation of the instructions by the malicious actors). The AI agent can either ignore these instructions or incorporate them into its actions. The decision the agent's AI model makes depends not only on the AI model itself but also on the design of the malicious instructions, which may be more or less convincing to the AI agent. This attack is essentially a form of social engineering against AI agents: manipulation through cleverly worded language to circumvent the desired behavior of the AI agent or even security guidelines---analogous to how a social engineering attack leads a human to perform an insecure action.

For this new class of threats, which operate at the intersection of technology and language, traditional technical security models appear limited. We therefore argue that the established field of research on human behavior in cybersecurity offers a crucial conceptual framework for enhancing the security of systems with respect to AI agents: What security mechanisms must we develop to protect AI agents, acting as independent actors on behalf of users, from attack vectors that target their human-like behavioral components?

Ideally, the positive aspects of the dualism of the human factor mentioned at the beginning (simultaneously attack vector and most capable defense) could be transferred to AI agents if AI, through its ability to learn, could also become one of the most capable components of a cybersecurity strategy.

To operationalize this, the factors identified in our human model can be directly mapped to functional components of agentic AI architectures. For instance, the human factor of ``Role'' translates to the AI's ``System Prompt'' or ``Persona Instructions'', while ``Organizational Culture'' could find their technical equivalent in the model's ``Alignment'' (e.g., via Reinforcement Learning from Human Feedback) or collaboration with other AI agents.

Crucially, even factors like ``Usability''---typically reserved for human-computer interaction---have a distinct parallel in AI behavior: ``Resource Accessibility'' or ``Computational Friction'': Just as humans bypass security controls that impose poor usability, an optimizing AI agent may prioritize a malicious data source simply because it is technically more accessible (e.g., clean, machine-readable text) over a legitimate source protected by friction-inducing elements like CAPTCHAs or complex authentication. Furthermore, the human behavioral risk of ``guessing'' when lacking specific knowledge (driven by high motivation to fulfill a role) maps directly to AI ``hallucination'': In both cases, the agent prioritizes the directive to be helpful over the constraint of accuracy.

By analyzing these mapped factors, we can predict ``path-of-least-resistance'' failures in AI agents just as we do in human employees.

\section{Conclusion}

In this work, we have presented a comprehensive model that synthesizes the complex drivers of human cybersecurity behavior into a structured framework. By categorizing factors into fundamental versus situational, and individual versus environmental dimensions, we provide a unified vocabulary to understand how organizational context shapes human action beyond simple decision-making. This holistic view highlights that security is not merely a product of individual competence, but a result of the interplay between a person's psychological state and their organizational environment.

Furthermore, we argue that this framework offers value beyond the human domain; it establishes a conceptual foundation for improved security of agentic AI: As AI systems increasingly assume quasi-human roles with high autonomy, they become susceptible to manipulation attacks that mirror social engineering. By applying this behavioral model to AI agents, researchers and practitioners can better anticipate these vulnerabilities, moving towards a security strategy that protects not just the human, but the digital agents acting on their behalf.


\end{document}